\documentclass[twocolumn,showpacs,preprintnumbers,amsmath,amssymb]{revtex4}
\usepackage{graphicx}% Include figure files
\usepackage{dcolumn}% Align table columns on decimal point
\usepackage{bm}% bold math

\begin{document}

\title{Studies of light nicleus clustering\\ 
in relativistic multifragmentation processes}% Force line breaks with \\

 \author{V.~Bradnova}
   \affiliation{Joint Insitute for Nuclear Research, Dubna, Russia} 
\author{M.~M.~Chernyavsky}
  \affiliation{Lebedev Institute of Physics, Russian Academy of Sciences, Moscow, Russia} 
 \author{A.~Sh.~Gaitinov}
   \affiliation{Institute for Physics and Technology, Almaty, Republic of Kazakhstan}
\author{L.~A.~Goncharova}
   \affiliation{Lebedev Institute of Physics, Russian Academy of Sciences, Moscow, Russia} 
\author{L.Just$^{\dag}$}
   \affiliation{P. J. \u Saf\u arik University, Ko\u sice, Slovak Republic} 
\author{S.~P.~Kharlamov}
   \affiliation{Lebedev Institute of Physics, Russian Academy of Sciences, Moscow, Russia}
\author{A.~D.~Kovalenko}
   \affiliation{Joint Insitute for Nuclear Research, Dubna, Russia}  
\author{M.~Haiduc}
   \affiliation{Institute of Space Sciences, Magurele, Romania}
\author{V.~G.~Larionova} 
  \affiliation {Lebedev Institute of Physics, Russian Academy of Sciences, Moscow, Russia}      
\author{F.~G.~Lepekhin}
  \affiliation{Petersburg Institute of Nuclear Physics, Gatchina, Russia}
\author{A.~I.~Malakhov}
   \affiliation{Joint Insitute for Nuclear Research, Dubna, Russia} 
\author{G.~I.~Orlova}
   \affiliation{Lebedev Institute of Physics, Russian Academy of Sciences, Moscow, Russia} 
\author{N.~G.~Peresadko}
   \affiliation{Lebedev Institute of Physics, Russian Academy of Sciences, Moscow, Russia}    
\author{N.~G.~Polukhina}
   \affiliation{Lebedev Institute of Physics, Russian Academy of Sciences, Moscow, Russia} 
\author{P.~A.~Rukoyatkin}
   \affiliation{Joint Insitute for Nuclear Research, Dubna, Russia} 
\author{V.~V.~Rusakova}
   \affiliation{Joint Insitute for Nuclear Research, Dubna, Russia} 
\author{N.~A.~Salmanova}
   \affiliation{Lebedev Institute of Physics, Russian Academy of Sciences, Moscow, Russia}    
\author{B.~B.~Simonov}
  \affiliation{Petersburg Institute of Nuclear Physics, Gatchina, Russia}
\author{S.~Vok\'al}
   \affiliation{P. J. \u Saf\u arik University, Ko\u sice, Slovak Republic}  
\author{P.~I.~Zarubin}
     \email{zarubin@lhe.jinr.ru}    
     \homepage{http://becquerel.lhe.jinr.ru}
   \affiliation{Joint Insitute for Nuclear Research, Dubna, Russia} 
 \author{I.~G.~Zarubina}
   \affiliation{Joint Insitute for Nuclear Research, Dubna, Russia}

\date{\today}% It is always \today, today,

\begin{abstract}
 \indent We give an overview of results and prospects of nuclear
clustering studies on the grounds of the  observations of interactions
 of light stable and radioactive nuclei with an initial energy above 1~A~GeV in nuclear emulsions.
Thank to the best spatial resolution and the full solid angle acceptance provided by nuclear emulsions, 
 such an approach allows one to obtain unique and evident observations reflecting  cluster-like 
 features in light nuclear structures.  New results on dissociation of 
$^7$Be in very peripheral interactions with emulsion nuclei are 
presented. The importance of this research for the physics of few body 
nuclear systems and the related problems of nucleosynthesis is noted. The paper is
illustrated with characteristic images obtained by means of
a microscope equipped with a CCD camera.	
The discussed explorations are provided with the beams of the Synchrophasotron 
and Nuclotron of JINR, Dubna. Future investigations are suggested to be carried out in relativistic beams 
of He, Be, B, C, and N  isotopes.\par
\end{abstract}
\maketitle

\section{\label{sec:level1}Introduction}
\indent The BECQUEREL Project (Beryllium (Boron) Clustering Quest in Relativistic 
Multifragmentation) is
 oriented toward emulsion expositions by light stable and radioactive 
nuclei with an energy of
 the order of few GeV per nucleon in the JINR Nuclotron beams. 
Observations of the fragmentation
of light relativistic nuclei open up new opportunities to explore highly 
excited nuclear states near
multiparticle decay thresholds \cite{Bradnova_FB}. Our interest in such 
states is motivated
by their predicted properties as loosely bound systems with spatial 
spread
significantly exceeding the fragment sizes. Natural components of such 
states are the lightest nuclei
 having no excited states below particle decay thresholds, i. e. 
deuterons, tritons, $^3$He, and $^4$He nuclei.\par

\indent A complete spectroscopy of few body decays of highly excited nuclei allows one to search for the Efimov 
excited states in nuclear systems \cite{Efimov,Nunes}. The Efimov effect is the remarkable theoretical observation
 that the number of bound states for three particles interacting via s-wave short range potentials 
 may grow to infinity, as the pair interaction are just about to bind two particles. 
The Efimov states are loosely bound and their wave functions extend far beyond those of the remaining 
two particles. Examples of such states in molecular physics are loosely bound helium molecules - 
dimmer H$_2$ (of a size about 50 \AA), the ground and the first excited state H$_3$. When the three particles are not identical, 
there are new possibilities for the existence of three-body ground states, like $^4$He-$^4$He-$^n$H (n=1-3)
 and $^4$He-$^4$He-$^{3}$He, He-He-Li, He-He-Na. Since $H$ and $He$ are the most abundant in the universe, 
 these molecules may play a role in the chemistry of the interstellar molecular cloud \cite{molec}.
Search for analogs of such states on nuclear scale is of undoubted interest since they can play a role of
intermediate states ("waiting stations") due to dramatically 
reduced Coulomb repulsion for nuclear fusions in stellar media.\par
\indent Other intriguing conjecture is that n$\alpha$ nuclei near the n 
$\alpha$ particle decay threshold can
constitute a loosely bound dilute gas forming a Bose condensate 
\cite{Ropke_CR}. Its major signature is a multiple
$\alpha$ particle decay with a narrowed distribution of relative 
velocities. Search for such states
on the nuclear scale is of undoubted interest since they can play a role 
of intermediate states ("waiting stations") for a stellar nuclear fusion due to dramatically 
reduced Coulomb repulsion.\par

\section{\label{sec:level2}Research concept}

\indent In this respect, a principal experimental task consists in provision of a complete spectroscopy
 of final fragments - observation of dissociation events, determination of various channel probabilities 
 (branchings), and fragment identification and velocity measurement. Our approach is grounded
  on spectroscopy of relativistic fragments of incoming nuclei at longitudinal exposures of
  emulsion layers.\par
\indent The track detection is performed in emulsion layers measuring
100$\times$200$\times$0.5 mm$^3$. The layers are assembled in few cm thick stacks. The stacks are 
exposed to a beam in the longitudinal direction. They provide multiple
track visualization over the total solid angle with spatial resolution of about 
0.5 micron. Their sensitivity extends from slow fragments down to relativistic single charged particles.
A mean range of light relativistic nuclei in emulsion is defined by the cross 
section of their inelastic interaction with emulsion nuclei. It varies from 14 cm for $^{6}$Li nuclei 
to 9 cm for $^{24}$Mg in a BR-2 type emulsion.\par 
   \indent The advantages with respect to low energy researches are the following:
\begin{itemize}

\item interactions reach a limiting fragmentation regime above a collision energy of 1A GeV and complex nuclear
 composition of emulsion doesn't affect isotopic composition of incoming nucleus fragments, 
 
\item reactions take the shortest time, especially in cases of electromagnetic and diffractive dissociations, 

\item projectile nucleus fragments are 
collimated mostly in a narrow forward cone limited by an angle 0.2/P$_0$, where P$_0$ is a 
primary nucleus momentum; this allows one to obtain 3D image of separated tracks in a single emulsion layer,

\item ionization losses of projectile fragments  correspond to the minimum and practically don't distort measurements,

\item the detection energy threshold for projectile fragments is absent,

\item a record spacial resolution of emulsion (0.5 micron) provides a record angular resolution;
the  excitation scale of a fragmenting system in the projectile rest frame 
is of the order of few MeV per fragmenting nucleon; in the case of the projectile nucleus fragmentation  
perfect angular measurements play the determinative role in the estimation of the excitation energy scale while their 
momentum per nucleon may be taken the same as for an incoming nucleus,
 
\item reliable determination of charges of relativistic fragments in provided over a wide range,
	
\item nuclear clustering manifests itself in the isotope composition of projectile nucleus fragments;
via measurement of a total momentum by multiple scattering technique one can identify hydrogen 
and helium isotopes;  

\item it is possible to select peripheral interactions corresponding to minimal energy transfers to an 
incoming nucleus; namely, these events present a major interest for studies of inΦlight multiparticle decays.

\end{itemize}
\indent
Visual scanning is concentrated on the events with a total 
charge transfer of an incoming nucleus to secondary particles in a 
narrow fragmentation cone. Emulsion nucleus fragmentation and meson 
production become reduced or even suppressed in this way. Such events 
amount to a few percent of the total number of inelastic events. In 
practice, this approach allows one to accumulate statistics of a few 
tens of peripheral events, which is sufficient for a reliable 
determination of dominating dissociation channels.
Thus, emulsions provide an excellent opportunity of a complete observation and 
study of light nucleus multifragmentation in flight. The Appendix contains a collection of 
characteristic images obtained by means of a microscope filming with a CCD camera
and reconstructed as flat projections.\par
\begin{figure*}
\begin{center}
\includegraphics[width=135mm]{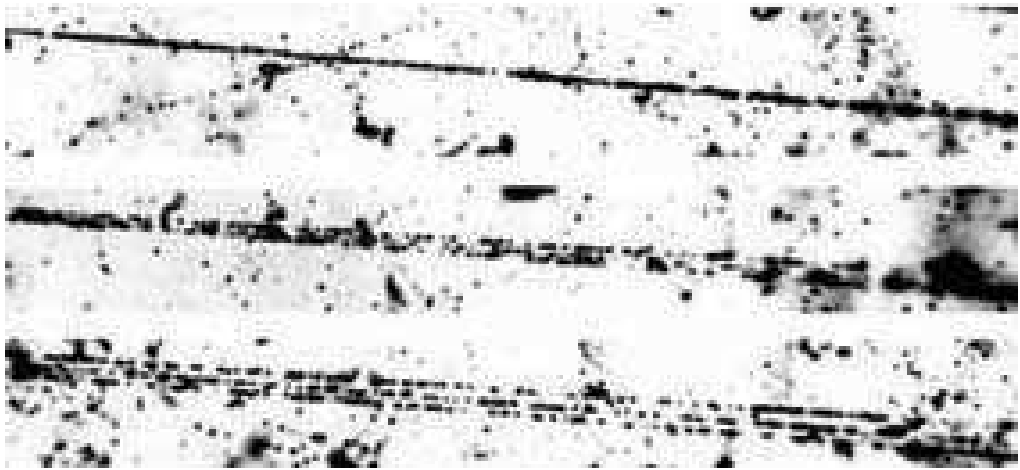}
\caption{\label{fig:1}Event of dissociation ofa  4.5~A~GeV/c $^{12}$C nucleus in peripheral interaction 
into three $\alpha$ particles shown in subsequent evaluation.
 Upper photo: interaction vertex with production of a narrow fragment jet.
 Middle and lower photo: shifting from vertex allow one to identify
  three double charged fragments.}
\end{center}
\end{figure*}
\begin{figure*}
\begin{center}
\includegraphics[width=135mm]{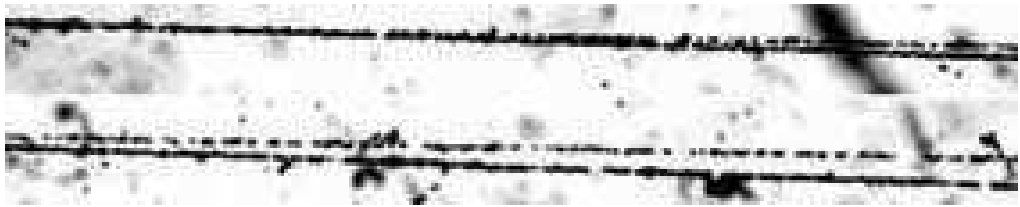} 
\caption{\label{fig:2}Event of asymmetric binary splitting
 of a 4.5~A~GeV/c $^{16}$O nucleus in peripheral interaction.
 Upper photo: interaction vertex with production of narrow fragment jet.
 Lower photo: shifting from vertex allows one to resolve relativistic
 C and He fragments.}
\end{center}
\end{figure*}

\begin{figure*}
\begin{center}
\includegraphics[width=135mm]{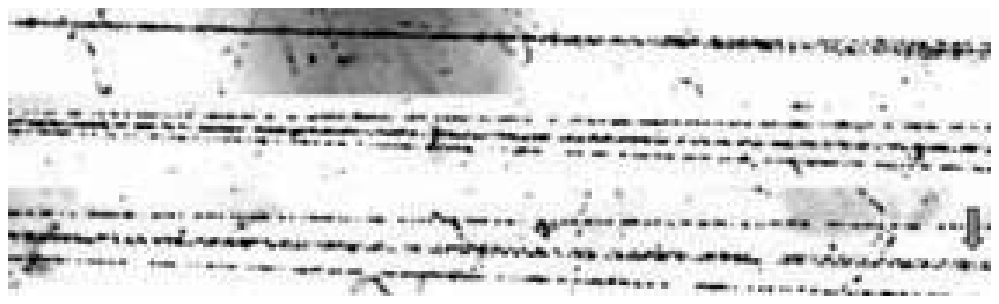}
\caption{\label{fig:3}Event of dissociation of a 4.5~A~GeV/c $^{16}$O nucleus 
in peripheral interaction into four He fragments.
 Upper photo: interaction vertex with production of a narrow fragment jet.
 Middle photo: shifting from vertex allow one to identify two He 
fragments and a very close track pair.
 Lower photo: further shifting allows one to resolve the central pair on the 
previous photo as  a relativistic $^8$Be production.}
\end{center}
\end{figure*}
\indent For the first time, a coherent multifragmentation was observed and explored in emulsion for $^{12}$C nucleus 
dissociation into three $\alpha$ particles at an initial momentum 4.5~A~GeV/c 
\cite{Clust01}-\cite{Clust03}. An example is shown in Fig.~\ref{fig:1}. The mean free path in emulsions for 
such events is equal to 5 m, i. e. this study demanded a visual scanning over a long track path to find the 
statistics of about 100 events. This scale of the statistics is a practical limit of the described approach. 
Relativistic multifragmentation was also explored for $^{16}$O (examples in Fig.~\ref{fig:2}~and~\ref{fig:3}), $^{24}$Mg, $^{28}$Si
 nuclei at an initial momentum 4.5~A~GeV/c by means of nuclear emulsion \cite{Clust04}-\cite{Clust07} and hydrogen bubble chamber techniques
\cite{Clust08}-\cite{Clust10}.\par
\indent Exotic nuclei present a bright extention of nuclear clustering in ground states. The lightest one among 
them is a $^{6}$Li nucleus. The important motivation for realizing the
 above mentioned feasible ways in the 
investigation of light nuclei became the results of the interactions of relativistic $^6$Li nuclei with 
emulsion nuclei \cite{Clust12}-\cite{Lepekhin_Li6}. We summarize below the results of this investigation basing 
mostly on the work \cite{Adamovich_Li6}. They serve to be a prototype of practical tasks for radioactive
 nuclei suggested in this paper.\par
 \indent At present an analogous analysis of $^{14}$N, $^7$Li and 
$^{11}$B exposures is carried out in order to find effects of deuteron 
and triton clusterings. Our first results for  $^7$Be and $^{22}$Ne nuclei
 are discussed in what follows.\par
 
\section{\label{sec:level3}Clustering in $^6$Li nucleus}
\indent A stack formed by emulsion layers was exposed to a beam of 4.5~A~GeV/c $^6$Li nuclei at the 
Synchrophasotron. During irradiation, the beam was directed in parallel to the emulsion plane. 
The first intriguing feature is that the mean free path of $^6$Li nuclei was found to be 
strongly decreased as compared with the expected value. The obtained value would  
correspond to a nucleus with mass number A equal to 11. This points to an unusually large radius 
of the nucleon distribution in $^6$Li nucleus. Using the geometric overlapping model, its value was 
estimated to be 2.7$\pm$0.1 fm, which is in a reasonable agreement with the known data on elastic 
scattering of protons on a $^6$Li target.\par
\indent Another distinctive feature of the $^6$Li nucleus was got by means of a multiple track 
scattering analysis, which allowed one to establish the isotopic composition of relativistic fragments. 
An unusually enhanced yield of relativistic deuterons was established to be the same as 
the proton one. A subsequent analysis dealt with $^3$He and $^4$He nuclei. The fragmentation of
 $^6$Li in the 
form of clusters consisting of $^3$He and tritium nuclei was shown to be by an order of magnitude 
weaker than the structure produced by an $\alpha$ particle and a deuteron. This explains an increased yield of 
deuterons as a reflection of the structure of weakly bound clusters of the $\alpha$ particle and the deuteron.
\par
\indent The fragmentation channel $^6$Li points to a lower value of the mean transverse
momentum of $\alpha$ particles,  $<$P$_T>$=0.13$\pm$0.1 GeV/c when compared with the case $^{12}$C having 
$<$P$_T>$=0.24$\pm$0.01 GeV/c. In the spirit of an uncertainty relation, this fact is another indication to an 
increased size of $^6$Li nucleus.\par
\begin{figure*}
\begin{center}
\includegraphics[width=135mm]{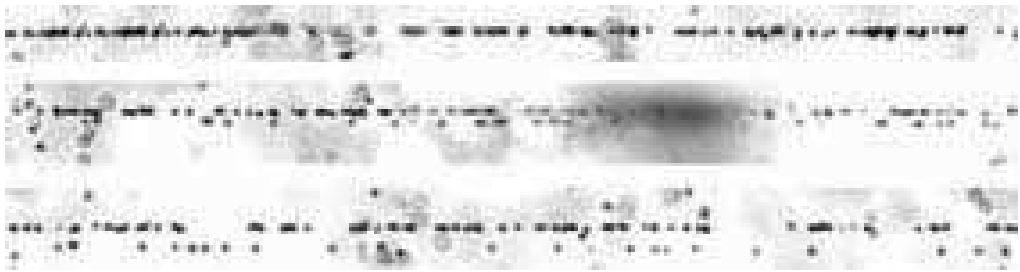}
\caption{\label{fig:4}Event of dissociation of a 4.5~A~GeV/c $^6$Li nucleus in peripheral interaction 
into H and He fragments.
 Upper photo: interaction vertex with production of a very narrow fragment pair.
 Lower photo: shifting from vertex allow one to resolve
 double and single charged relativistic fragments.}
\end{center}
\end{figure*}

\indent Among the 1000 found $^6$Li interactions it is possible to consider as "golden" 31 events of 
coherent $^6$Li dissociation not accompanied by the target nucleus excitation ("white stars").
An example is shown in Fig.~\ref{fig:4}. Among them 
23 events correspond to the dissociation channel $\alpha$+d, four of them to $^3$He+t, four to t+d+p and 
none of them to d+d+d. This topology shows the cluster structure of $^6$Li nucleus in the most obvious 
manner. Thanks to the completely reconstructed coherent dissociation kinematics it became 
possible to reconstruct the $^6$Li levels at 2.19 and 4.31 MeV with isotopic spin T=0. On the contrary, 
the 3.56 MeV level with isotopic spin T=1 is absent because of the $\alpha$+d system isotopic spin T=0. 
This is the very clear illustration of an isotopic filtering in strong interactions.\par

\section{\label{sec:level4}Clustering in $^{10}$B dissociation}
\indent $^{10}$B nuclei were accelerated at the JINR Nuclotron, and a $^{10}$B nucleus beam of energy
  of 1A GeV was formed. The beam was used to irradiate stacks composed of BR-2 type emulsion 
  layers. During irradiation, the emulsion layers were 
  located in parallel to the beam direction so that the beam particles could enter the 
  butt-end of the emulsion layers. Search for nucleus-nucleus interactions was performed 
  by visual scanning of particle tracks by means of microscopes ($\times$ 900 magnification). 
  At a scanned track length of 138.1 m, it was found 960 inelastic interactions of   $^{10}$B   nuclei. 
  The mean free path of $^{10}$B nuclei to an inelastic interaction in emulsion was found to be 
  14.4$\pm$0.5 cm. This value meets well the dependence of the mean free path upon the atomic number
   of a bombarding nucleus for light nuclei having homogeneous   nucleonic density.\par

 \begin{figure*}
\includegraphics[width=135mm]{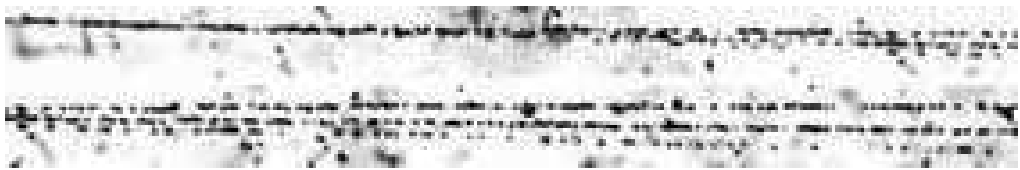}
\caption{\label{fig:5}Event of dissociation of a 1~A~GeV $^{10}$B nucleus into two 
double and one single charged fragments.}
\end{figure*}  
 \indent Information about the charge composition of charged fragments and about the channels of $^{10}$B 
nucleus fragmentation in peripheral collisions has been obtained. We attribute to the peripheral 
interactions events in which the total charge of relativistic fragments is equal to the charge of 
the primary $^{10}$B nucleus, the production of charged mesons is not observed, while the production of 
slow nuclear   fragments is allowed in selection. In order to single out such events by visual observation we
 estimated the charges of relativistic particles and the total charge of relativistic particles
  with emission angles less than 15$^o$ with respect to the $^{10}$B direction. For the primary beam 
  energy of 1A GeV, this value corresponds to the proton transverse momentum of 0.44~GeV/c.
   Then using measuring microscopes we evaluated the emission angles of all particles in the 
   selected events. The particle charges were determined by the length   of spacings in their 
   tracks. \par
  \indent The number of the found events in which the total charge of fragments is equal to 
   five and in which charged mesons are not observed is equal to 93 (10\% of all the events). 
   We notice that the selection of events in which the production of even nuclear emulsion 
   fragments is forbidden decreases statistics down to 41 events ("white stars"). In this case, the distribution 
   of statistics by channels remains practically unchanged.\par

\begin{figure*}
\includegraphics[width=135mm]{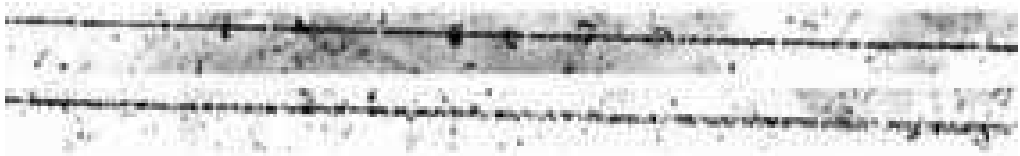}
\caption{\label{fig:6}Event of a 1A GeV$^{10}$B nucleus dissociation to a
Li (top) and He fragments (bottom).}
\end{figure*}

   \indent In 65\% of such peripheral interactions the $^{10}$B nucleus is disintegrated to two
 double charged and a one single-charged particles (Fig.\ref{fig:5}). A single-charged particle 
 is the deuteron in 40\% of these events. 10\% of events contain fragments with
  a charge equal to 3 and 2 (Li and He isotopes), and 2\% of events contain a fragment
   with charges equal to 4 and 1 (Be nucleus and the proton). The $^6$Li production accompanied   
   by an alpha particle may be considered as a correlation of $\alpha$ particle and deuteron clusters.
    Fig.~\ref{fig:6} shows an example of a two-particle decay to Li and He fragments.
     The fragmentation channel containing $\alpha$ particle and three single-charged fragments 
     (disintegration of one of the $\alpha$ clusters) makes up 15\%.
An equal correlation of the channels (He+He+d)/(He+He+p)$\approx$1 is analogous to the $^{10}$Li
 fragmentation where (He+d)/(He+p)$\approx$1. These ratios point to an abundant yield of
  deuterons in the $^{10}$B case too \cite{Adamovich_Li6,Lepekhin_Li6}.\par
  \indent Thus, the deuteron cluster manifests itself directly 
  in the three-particle decays of $^{10}$B nuclei accompanied by two two-charged particles.
   Another indication to deuteron clustering is a small mean transverse momentum of 
   deuterons P$_T$=0.14$\pm$0.01~GeV/c in these events, in just the same way as in the case of 
   $^6$Li, where P$_T$= 0.13$\pm$0.02~GeV/c.\par
\begin{figure*}
\includegraphics[width=135mm]{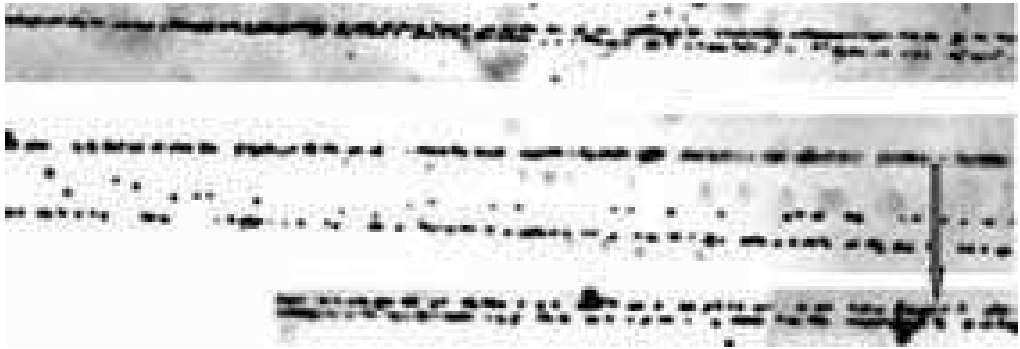}
\caption{\label{fig:7}Event of dissociation of a 4.5~A~GeV/c $^{14}$N nucleus 
in peripheral interaction into three He and one H fragments.
 Upper photo: interaction vertex with production of a narrow fragment jet.
 Middle photo: shifting from vertex allows one to identify single H 
 and He relativistic fragments accompanied by a very close track pair (top).
 Lower photo: further shifting allows one to resolve the central pair on the 
previous photo as  a relativistic $^8$Be production.}
\end{figure*}

   \indent
	We note that $^{10}$B nucleus, like the deuteron, and $^6$Li and $^{14}$N belong to a rare
class of even-even stable nuclei. Therefore, it is interesting to establish the presence of deuteron
 clustering in relativistic $^{14}$N fragmentation as well as 
the appearance of $^8$Be clustering (example in Fig.~\ref{fig:7}).\par
	
\section{\label{sec:level5}Clustering in $^7$Li dissociation}
\indent A total of 1274 inelastic interactions were found to be occurred 
in a nuclear emulsion irradiated by a $^7$Li beam with a momentum 3~A~GeV/c at the JINR Synchrophasotron,
 at a length of 185 m of scanned tracks. The mean free path of   $^7$Li nuclei up to an inelastic 
 interaction in emulsion was found to be 14.5$\pm$0.4 cm which coincides, within errors, with 
 the $^7$Li mean free path \cite{Clust12,Adamovich_Li6,Lepekhin_Li6}.
  The close values of the mean free paths and the total transverse 
 cross sections of inelastic interactions of $^6$Li and $^7$Li point out that their effective interaction 
 radii are also close in magnitude to each other.\par

 \indent About 7\% of all inelastic interactions of $^7$Li nuclei are peripheral interactions (80 events),
	 which contain only the charged fragments of a relativistic nucleus, they do not contain 
	 any other secondary charged particles, and the total charge of the fragments is equal to
	  the charge of a fragmenting nucleus ("white stars"). All these events are actually two-particle $^7$Li 
	  decays to one double and one single-charged fragments. Half of these events are attributed 
	  to a decay of $^7$Li nucleus to an $\alpha$ particle and a triton (40 events). The number of decays 
	  accompanied by deuterons makes up 30\%, and by protons - 20\%. \par
\indent The isotopic composition 
	  of decayed particles points to the fact that these events are related to the structure 
	  as the $\alpha$ particle and triton clusters. The predominance of tritons in the isotopic compound
	   of single-charged fragments shows well the dominating role of the triton cluster in the $^7$Li
	    fragmentation in very peripheral interactions with emulsion nuclei.\par
	    \indent Similar two-particle $^6$Li decays to an $\alpha$ particle and a deuteron which reflected the weakly
 bound two-cluster nuclear structure were registered in inelastic peripheral interactions of
  $^7$Li nuclei with a momentum 4.5~A~GeV/c in emulsion. 
  Thus, the structure in the form of the $\alpha$ particle core and external nucleons bound into a cluster
   is typical not only of $^6$Li nucleus, but also of $^7$Li one. The obtained value of the cross section 
   for coherent decay to an $\alpha$ particle and a triton, 27$\pm$4 mb, was found to be about the same
    as the cross section of paper \cite{Adamovich_Li6} for $^6$Li decay to $\alpha$ particle and a deuteron 22$\pm$4 mb.
     This may be viewed as an indication to the fact that the mechanisms of the decays 
     in question are of the same nature. It is interesting to continue to clear up a possible
      role played by the tritons as cluster elements in $^{11}$B and then in $^{15}$N nuclei.\par

\section{\label{sec:level6}Clustering in $^6$He dissociation} 

\indent Using a $^6$Li nucleus beam extracted from the JINR Synchrophasotron at momentum 2.67~A~GeV/c
 a secondary beam was produced with a composition of 1\% of $^6$He and 99\% of $^3$H nuclei \cite{Clust17}.\par
\begin{figure*}
\includegraphics[width=135mm]{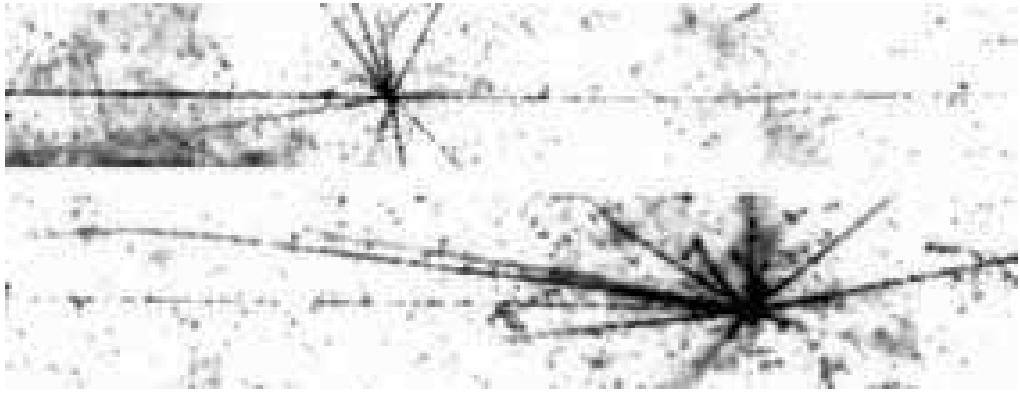}
\caption{\label{fig:8}Event of a 2.67~A~GeV $^6$He nucleus interaction with fragmentation 
into a relativistic $\alpha$ particle. The $\alpha$ particle 
track is followed until a violent inelastic interaction.}
\end{figure*}

\indent The major feature of $^6$He nucleus interactions with nuclei is due to a peculiar interaction of 
an external neutron pair with the target nucleus. The process $^6$He$\rightarrow\alpha$+n+n
 is possible in a peripheral interaction 
in which a break-up of external neutrons occurs without a secondary particle production ("white stars"). 
  An example is shown in Fig.~\ref{fig:8}.The 
determination of the process probability and measurement of the angular distribution of produced $\alpha$ 
particles is one of the major tasks of $^6$He nucleus experimental studies. Such events are presented in 
emulsion by an image of a scattered track of a double charged particle. Analogous $^6$Li nucleus 
interactions in emulsion with $\alpha$ particles in the final state were observed 
due to reactions $^6$Li$\rightarrow\alpha$+d and $^6$Li$\rightarrow\alpha$+p+n
 in \cite{Adamovich_Li6}.\par
 \indent Indeed, in this study, on a 457 cm scanned path of double charged tracks it was 
found 23 inelastic interactions having secondary particles  and, in addition, 
5 events of a gradual change of the direction behind a scattering point. The mean momentum in 
these five events is equal to 15.6$\pm$3.8~GeV/c, and after scattering to 9.1$\pm$2.6~GeV/c. The scattering 
angle do not exceed 0.35$^o$ in all the found cases, and the mean transverse momentum of $\alpha$
particles $<$P$_T>$ is about 0.035~GeV/c. We notice that in dissociation processes 
$^6$Li$\rightarrow\alpha$+d and $^6$Li$\rightarrow\alpha$+p+n at 
a $^6$Li nucleus momentum 4.5~A~GeV/c the corresponding $<$P$_T>$ is about 0.15 
GeV/c \cite{Clust12,Adamovich_Li6,Lepekhin_Li6}.\par
\indent A narrower transverse momentum distribution in the process points to a very 
peripheral picture of $\alpha$ particle production in a coherent interaction.
A mean free range defined on the 457 cm path taking into account the registered coherent 
interactions is 16.3$\pm$3.1 cm. This value is larger than the corresponding one for
 a $^6$Li 
nucleus determined in \cite{Clust12,Adamovich_Li6,Lepekhin_Li6} as 14.3$\pm$0.3 cm.
 It might be assumed that the excessive value
can be explained by an assumption of a 50\% identification efficiency of the process. The 
validity of such an assumption will indicate that the contribution of coherent interactions to the 
total cross-section exceeds 20\%.\par

\section{\label{sec:level7}Clustering in $^7$Be dissociation}

\indent Advantages of emulsion technique are exploited most completely in the 
study of peripheral fragmentation of light stable and neutron deficient 
nuclei.
 A secondary beam containing a significant fraction of 1.23~A~GeV $^7$Be 
nuclei was formed at the JINR Nuclotron by selecting the products of 
charge exchange of primary $^7$Li nuclei with the aid a beam 
transport channel.
Emulsion stacks have been irradiated. The 
$^7$Be nucleus is convenient for magnet optics selection due to the 
maximum ratio of the charge to the weight. Besides, it gives the most 
complete observation of final fragments.\par
\begin{figure*}
\includegraphics[width=135mm]{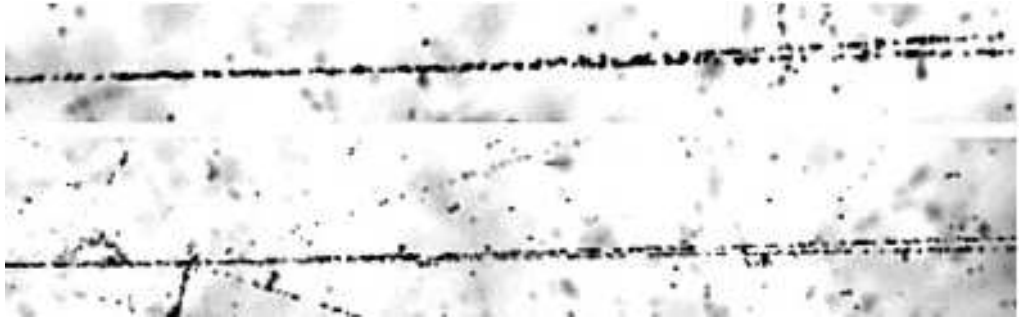}
\caption{\label{fig:9}Examples of peripheral dissociation of 1.23~A~GeV $^7$Be 
nuclei into pairs of He nuclei.
 Upper photo: dissociation without target nucleus excitation and
  produced charged mesons. Lower photo: dissociation accompanied
   by a target fragment and a meson like pair.}
\end{figure*}

\indent  By visual scanning along tracks, we have found 22 decays of incoming 
nuclei to helium fragments without other accompanying tracks ("white stars"). The event 
examples are shown in
Fig.~\ref{fig:9}. Helium isotopes have been identified via their total momenta 
derived from multiple scattering measurements. This makes it possible to conclude, that 
a dominant fraction is due to a coherent dissociation
$^3$He+$^4$He and only 2-3 decays to $^3$He+$^3$He+n.
We have found 20 events with charged topology of relativistic 
fragments 2+1+1 ("white stars") with identified $^3$He and 16 with identified $^4$He.
The events with topology 3+1 and 1+1+1+1 are presently analysed. They represent only a few 
percent fraction. \par
\indent Thus, one can conclude that a $^3$He clustering manifests in decays of excited 
relativistic $^7$Be nuclei.
 This result agrees with our expectations. It seems to be useful for future 
studies of the role of $^3$He clusterization in 
three body decays of $^8$B ($^{1,2}$H-$^{4,3}$He-$^3$He), $^9$C 
($^3$He-$^3$He-$^3$He),
$^{10}$C ($^3$He-$^3$He-$^4$He), and $^{11}$C ($^3$He-$^4$He-$^4$He).
It is quite possible that $^3$He clustering could play a role analogous 
to the role of a triple $\alpha$ process in stellar nucleosynthesis.\par

\section{\label{sec:level8}Clustering in $^{22}$Ne dissociation}

\indent  We reanalyzed of 
existing data (DST) on about 4100 $^{22}$Ne 4.1~A~GeV/c interactions in an emulsion.
We found 94 events containing only fragments of a primary nucleus 
without target nucleus fragments and produced mesons ("white stars").\par
\begin{figure*}
\includegraphics[width=135mm]{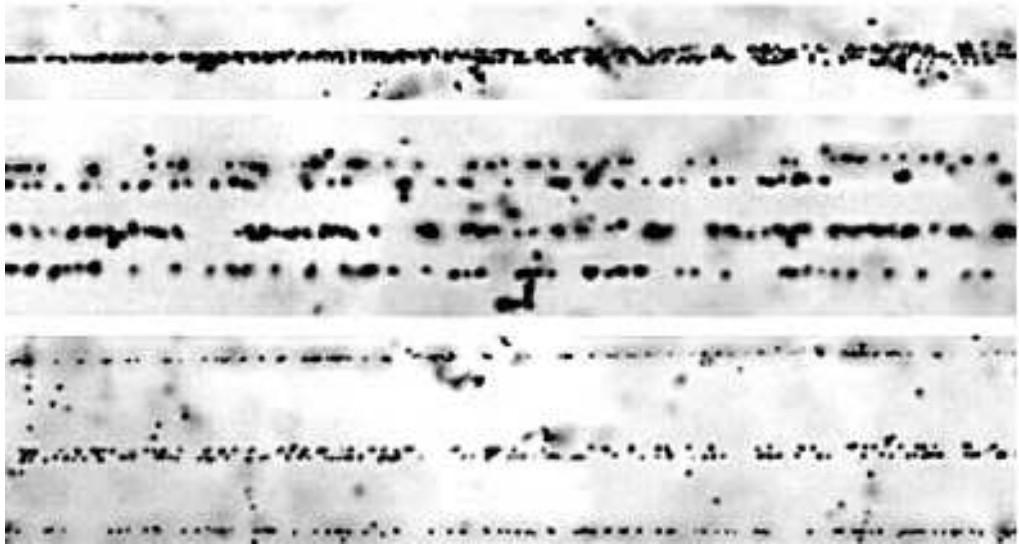}
\caption{\label{fig:10}Dissociation of a 4.5~A~GeV/c $^{20}$Ne nucleus in peripheral interaction 
into five He fragments.
 Upper photo: Interaction vertex with production of a narrow fragment jet.
 Middle photo: shifting from vertex allows one to identify three He 
fragments.
 Lower photo: further shifting allows one to resolve a central track on the 
previous photo as a very
narrow pair of relativistic He nuclei ($^8$Be production).}
\end{figure*}

\indent The dominant channel is 8+2  $-$ 53 events (He cluster separation), the 
follow 9 + 1 $-$ 14, 7 + 2 + 1 $-$ 7, 8 + 1 + 1 $-$  6, 6 + 2 + 2 $-$ 5, 6 + 2 + 1 + 1 $-$  
3. Among deeper fragmentation there are 2 events 4+1+1+1+1+1+1 and
4 + 2 + 2 + 2, and single events with topologies  5 + 2 + 1+ 1+ 1, 5 + 2 + 2 + 1, 3 + 2 + 1 + 1 + 1 + 1 + 1,  
and 2 + 2 + 2 + 2 + 1 + 1. The most interesting fact is that we have 
found 3 events of topology 2+2+2+2+2, corresponding neon dissociation 
into helium nucleus fragments only. The example is shown in the
Fig.~\ref{fig:10}.\par
\indent In our opinion, such distribution of final charged states brightly 
illustrate transition from a single helium fragment splitting to a 
total multifragmentation of the explored nucleus. Absence in the 
selected statistics of binary fissions like 7+3, 6+4, 5+5 is especially 
interesting. We plan to carry out a detailed kinematical analysis of these 
events.\par
\begin{figure*}
\includegraphics[width=135mm]{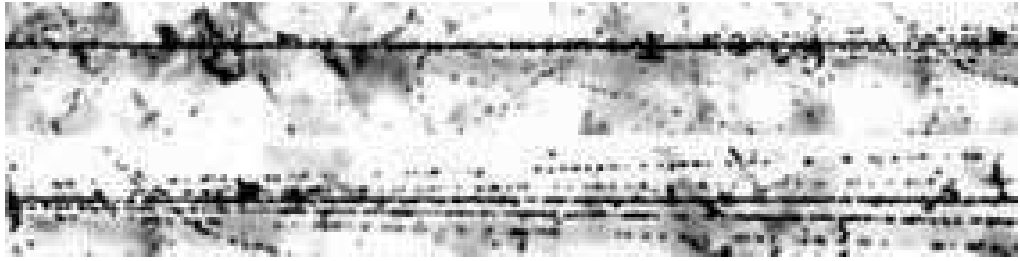}
\caption{\label{fig:11}Event of dissociation of 4.5~A~GeV/c $^{24}$Mg 
nucleus in peripheral interaction 
into charged topology 1+1+7+2+1 (from top to bootom).
 Upper photo: interaction vertex with production of narrow fragment jet 
 accompanied with  a relativistic meson like track.
Lower photo: shifting from vertex allows one to resolve five 
fragments.}
\end{figure*}
\begin{figure*}
\includegraphics[width=135mm]{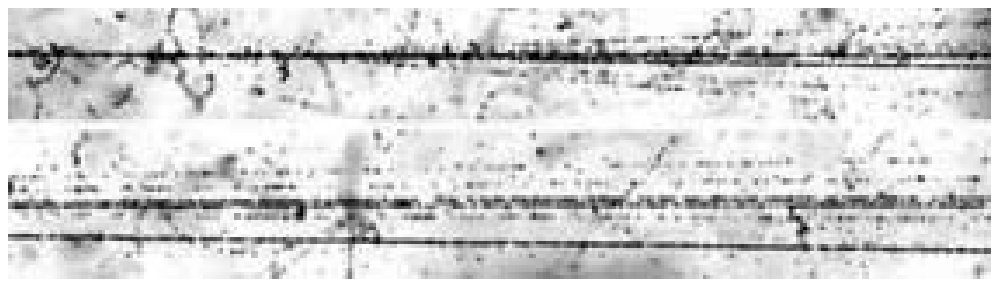}
\caption{\label{fig:12}Event of dissociation of a 4.5~A~GeV/c $^{24}$Mg 
nucleus in peripheral interaction 
into charged topology 1+1+2+2+1+5 (from top to bootom).
 Upper photo: interaction vertex with production of a narrow fragment jet 
 accompanied with a relativistic meson like track.
Lower photo: shifting from vertex allows one to resolve six 
relativistic fragments.}
\end{figure*}
\begin{figure*}
\includegraphics[width=135mm]{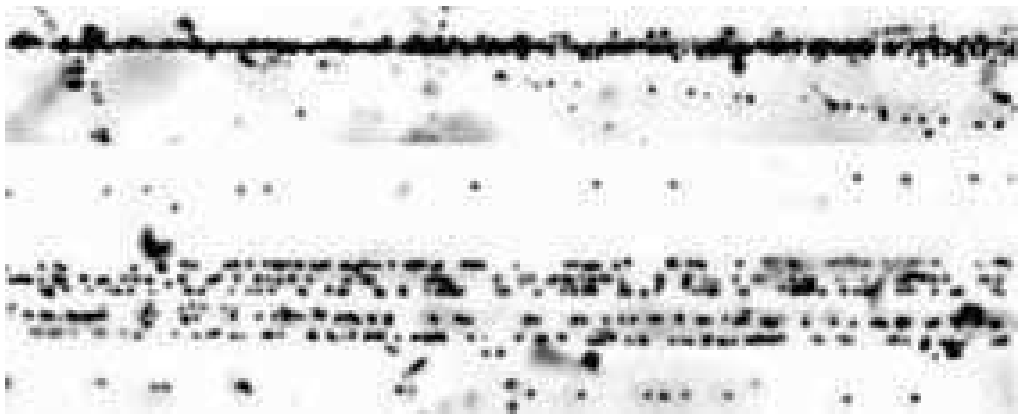}
\caption{\label{fig:13}Event of a 4.5~A~GeV/c $^{24}$Mg nucleus dissociation into 
five double and two single charged
fragments.
 Upper photo: interaction vertex with production of narrow fragment jet 
 accompanied with a relativistic meson like track.
Lower photo: shifting from vertex allows one to resolve seven 
relativistic fragments.}
\end{figure*}
\begin{figure*}
\includegraphics[width=135mm]{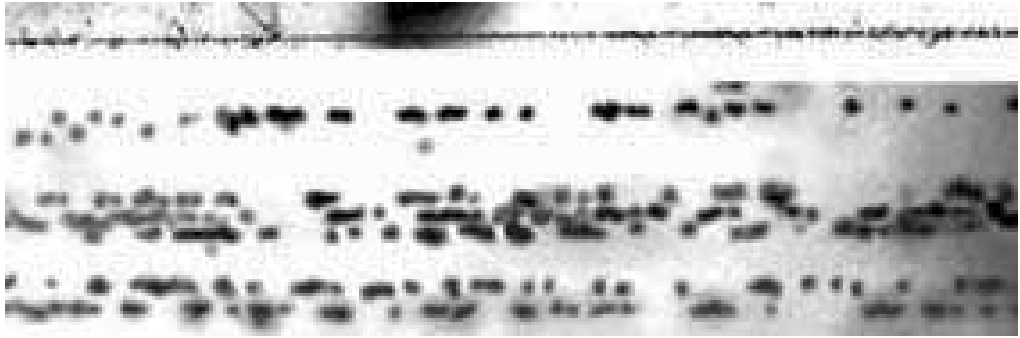}
\caption{\label{fig:14}Event of dissociation of a 4.5~A~GeV/c $^{24}$Mg nucleus 
in peripheral interaction into six He fragments.
 Upper photo: interaction vertex with production of a narrow fragment
  jet accompanied with a 
 couple of target slow fragments.
 Lower photo: further shifting allows one to resolve (from top to bottom)
  separate He fragment,
 very narrow He fragment triple (most probably, decay of excited 
 $^{12}$C nucleus), and a very narrow He pair ($^8$Be production).}
\end{figure*}
\begin{figure*}
\includegraphics[width=135mm]{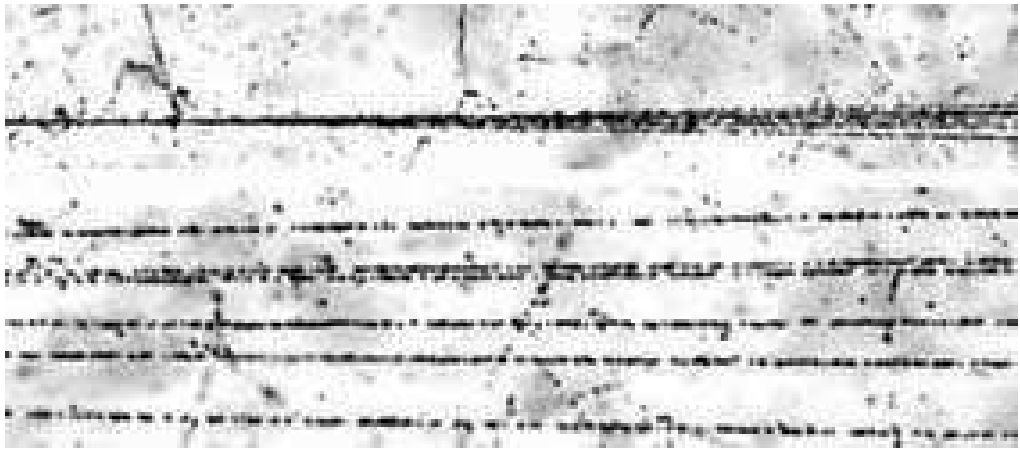}
\caption{\label{fig:15}Event of dissociation of a 4.5~A~GeV/c $^{24}$Mg nucleus 
in peripheral interaction into five $\alpha$ particles and single 
$^3$He fragments.
 Upper photo: interaction vertex with production of narrow fragment
  jet accompainied with a target recoil.
 Lower photo: further shifting allows one to resolve (from top to bottom)
 a six He fragment system containing a very narrow $\alpha$ particle
   pair ($^8$Be production).}
\end{figure*}

\indent Universality of a coherent dissociation mechanism enables us to search 
for such events in emulsions irradiated by $^{24}$Mg, 
$^{28}$Si, and $^{32}$S nuclei at 4.5~A~GeV/c in order to study a 
relative role of multiparticle decays. Examples of $^{24}$Mg peripheral 
interactions with growing dissosiation degree are shown in the Fig.~\ref{fig:11}~-~\ref{fig:15}.\par
\indent Verification of a hypothesis about the phase transition of light nuclei 
from a ground state to multiparticle one via a Bose condensate is one of 
intriguing perspectives of this research.\par

\begin{acknowledgments}
\indent The development of our project would not have been possible without
 scientific guidance and support of late Academician A. M. Baldin.
  We are indebted to late Prof. M. I. Adamovich for many year leadership
   in emulsion technique. Papers of late Prof. G. M. Chernov on nuclear 
   coherent dissociation has played inspiring role in our research. Our 
   manuscript is devoted to the memory of these outstanding scientists.\par
\indent The authors express their warmest thanks to A. V. Pisetskaya from FIAN, 
Moscow and I. I. Sosulnikova, A. M. Sosulnikova, N. A. Kachalova,  
G. V. Stelmakh, and A. Vok\'alova from JINR, Dubna for their contribution in a microscope
 visual analysis and I. I. Marin for perfect maintenance of microscopes. 
 Emulsion proceeding was performed by the LHE JINR chemist group 
 with excellent quality.\par
\indent Valuable contributions to our work have been provided by the JINR
   Synchrophasotron and Nuclotron personnel and the beam transport group.\par 
\indent This work has
 been supported by the grants 96-15-96423,
03-02-16134, 02-02-164-12a of the Russian foundation of basic 
researches,VEGA N1/9036/02 of the Agency of Science of the Ministry of Education
of the Slovak Republic and the Slovak Academy of Sciences, and the 
grants of the JINR Plenipotentiaries of Slovakia, Czechia
and Romania for the years 2002 and 2003.\par
\end{acknowledgments}

\end{document}